\documentstyle[12pt,fleqn]{article}

\def\thebibliography#1{\section*{References}\list
  {[\arabic{enumi}]}{\settowidth\labelwidth{#1}\leftmargin\labelwidth
    \advance\leftmargin\labelsep
    \usecounter{enumi}}
    \def\newblock{\hskip .11em plus .33em minus .07em}
    \sloppy\clubpenalty4000\widowpenalty4000
    \sfcode`\.=1000\relax}

\def\op#1{\mathop{\fam0 #1}\limits}

\newcommand{\im}{{\rm Im\,}}

\newcommand{\id}{{\rm Id\,}}

\newcommand{\beq}{\begin{equation}}
\newcommand{\eeq}{\end{equation}}
\newcommand{\ben}{\begin{eqnarray}}
\newcommand{\een}{\end{eqnarray}}
\newcommand{\be}{\begin{eqnarray*}}
\newcommand{\ee}{\end{eqnarray*}}
\newcommand{\bea}{\begin{eqalph}}
\newcommand{\eea}{\end{eqalph}}

\newcommand{\cA}{{\cal A}}

\newcommand{\cD}{{\cal D}}

\newcommand{\cL}{{\cal L}}

\newcommand{\cH}{{\cal H}}

\newcommand{\bL}{{\bf L}}
\newcommand{\bR}{{\bf R}}
\newcommand{\bC}{{\bf C}}
\newcommand{\bZ}{{\bf Z}}

\newcommand{\bH}{{\bf H}}
\newcommand{\bT}{{\bf T}}

\newcommand{\al}{\alpha}
\newcommand{\vr}{\varrho}
\newcommand{\bt}{\beta}

\newcommand{\la}{\lambda}

\newcommand{\f}{\phi}
\newcommand{\om}{\omega}
\newcommand{\Om}{\Omega}
\newcommand{\m}{\mu}
\newcommand{\n}{\nu}
\newcommand{\g}{\gamma}

\newcommand{\up}{\upsilon}
\newcommand{\lng}{\langle}
\newcommand{\rng}{\rangle}

\newcommand{\w}{\wedge}
\newcommand{\wt}{\widetilde}
\newcommand{\wh}{\widehat}
\newcommand{\ol}{\overline}
\newcommand{\dr}{\partial}

\newcommand{\ot}{\otimes}

\newcounter{eqalph}
\newcounter{equationa}
\newcounter{theorem}
\newcounter{remark}
\newcounter{example}
\newcounter{proposition}
\newcounter{lemma}
\newcounter{corollary}
\newcounter{definition}

\newenvironment{eqalph}{\stepcounter{equation}
\setcounter{equationa}{\value{equation}}
\setcounter{equation}{0}

\begin{eqnarray}}{\end{eqnarray}\setcounter{equation}{\value{equationa}}}

\def\theremark{\arabic{remark}}

\def\thedefinition{\arabic{definition}}

\newenvironment{rem}{\refstepcounter{remark}\bigskip\noindent{\it
Remark \theremark.}}{\medskip}
\newenvironment{ex}{\refstepcounter{remark}\bigskip\noindent{\it
Example \theremark.}}{\medskip}

\newcommand{\mar}[1]{}

\begin{document}
\hbox{}

{\parindent=0pt

{\large\bf Geometric quantization of non-relativistic and
relativistic
 Hamiltonian mechanics}
\bigskip

{\sc Giovanni Giachetta$^\dagger$\footnote{E-mail
address: mangiaro@camserv.unicam.it}, Luigi
Mangiarotti$^\dagger$\footnote{E-mail address: mangiaro@camserv.unicam.it} and
Gennadi  Sardanashvily$^\ddagger$\footnote{E-mail address:
sard@grav.phys.msu.su}}
\medskip

\begin{small}
$\dagger$ Department of Mathematics and Physics, University of Camerino, 62032
Camerino (MC), Italy \\
$\ddagger$ Department of Theoretical Physics, Physics Faculty, Moscow State
University, 117234 Moscow, Russia

{\bf Abstract.}
We show that
non-relativistic and relativistic mechanical systems on a configuration
space $Q$ can be seen as the conservative Dirac constraint
systems with zero Hamiltonians on different
subbundles of the same cotangent bundle $T^*Q$. The geometric quantization
of this cotangent bundle under the vertical polarization leads to compatible
covariant quantizations of non-relativistic and relativistic
Hamiltonian mechanics.   
 
\end{small}
}


\section{Introduction}

We study
covariant geometric quantization of Hamiltonian mechanics subject to
time-dependent transformations, whose configuration space
is an
$(m+1)$-dimensional oriented smooth manifold $Q$ coordinated by $(q^\la)$.
This is the case of non-relativistic mechanics on a configuration space and
relativistic mechanics on a pseudo-Riemannian manifold.

In non-relativistic
mechanics,
$Q$ is a fibre bundle over the time axis $\bR$ provided with the Cartesian
coordinate $q^0=t$ with affine transition functions $t'=t+$const. 
Its different trivializations
$Q\cong \bR\times M$ 
correspond to different non-relativistic reference frames. 
In
relativistic mechanics, a configuration manifold $Q$ need no fibration over
$\bR$, but is a time-like oriented pseudo-Riemannian manifold with respect to a
non-degenerate metric $g$ of signature
$(+,-\cdots)$. It admits a coordinate atlas
$\{(U;q^0,q^k)\}$ where
$g^{00}>0$ on each coordinate chart. Such a coordinate chart
provides the local fibration 
\mar{gm600}\beq
U\ni (q^0,q^k)\mapsto q^0\in\bR, \label{gm600}
\eeq
and can be seen as a
local non-relativistic configuration space. In particular, if $Q=\bR^4$ and $g$
is the Minkowski metric, one comes to Special Relativity.

Let $T^*Q$ be the cotangent bundle $T^*Q$ of $Q$. Coordinated by
$(q^\la,p_\la=\dot q_\la)$, it is provided with the canonical Liouville
form
$\Xi=p_\la dq^\la$, the above mentioned symplectic form
$\Om=d\Xi$, and the corresponding Poisson bracket  
\be
\{f,g\}_T =\dr^\la f\dr_\la f -\dr_\la
f\dr^\la f' 
\ee
on the ring $C^\infty(T^*Q)$ of smooth real functions on $T^*Q$.
We will show that
the cotangent bundle $T^*Q$ of $Q$  plays a
role both of the homogeneous momentum phase space of non-relativistic
mechanics and the momentum phase space of relativistic mechanics on
$Q$, but non-relativistic and relativistic Hamiltonian
systems occupy different one-codimensional subbundles $N$ and $N'$ of $T^*Q\to
Q$. They are given by the constraints
\mar{qq90}\beq
(a)\,\, \f_N=p_0 +\cH(q^\la,p_k)=0, \qquad
(b)\,\, \f_{N'}=g_{\m\nu}\dr^\m\cH'\dr^\nu\cH'-1=0, \label{qq90}
\eeq
where $\cH$ and
$\cH'$ are non-relativistic and relativistic Hamiltonians, respectively. 
Solutions of
non-relativistic and relativistic Hamiltonian systems are vector fields on
$N$ and $N'$, which obey the constraint Hamilton equations 
\mar{qq91}\beq
(a)\,\, \g\rfloor\Om_N=0, \qquad (b)\,\, \g'\rfloor\Om_{N'}=0, \label{qq91}
\eeq
where $\Om_N$ and $\Om_{N'}$ are the pull-back presymplectic forms on $N$
and $N'$. These are equations of the conservative Dirac constraint dynamics
 on the symplectic manifold $(T^*Q,\Om)$, whose Hamiltonian is constant on
a primary constraint space. 

Therefore, in order to quantize non-relativistic and relativistic mechanics,
one can provide the geometric quantization of the cotangent bundle
$T^*Q$, where non-relativistic and relativistic
Hamiltonian systems are characterized by quantum constraints 
\mar{qq92}\beq
(a)\,\,\wh \f_N\psi=0, \qquad (b)\,\, \wh \f_{N'}\psi'=0 \label{qq92}
\eeq
on elements $\psi$ of the quantum space.

Recall that  the
geometric quantization procedure falls into the following three
steps: prequantization, polarization and metaplectic correction (e.g.,
\cite{eche98,sni,wood}).  Given a symplectic  manifold $(Z,\Om)$ and the
corresponding Poisson bracket $\{,\}$, prequantization associates to each
element $f$ of the Poisson algebra
$C^\infty(Z)$ on $Z$ 
a first order differential operator
$\wh f$ in the space of sections of a complex line bundle $C$ over $Z$
such that the Dirac condition
\mar{qm514}\beq
[\wh f,\wh f']=-i\wh{\{f,f'\}} \label{qm514}
\eeq
holds. 
Polarization of a symplectic manifold $(Z,\Om)$
is defined as a maximal involutive distribution $\bT\subset TZ$ such that 
Orth$_\Om \bT=\bT$, i.e., 
\mar{qq26}\beq
\Om(u,\up)=0, \qquad \forall u,\up\in\bT_z, \qquad z\in Z. \label{qq26}
\eeq
Given the Lie algebra $\bT(Z)$ of global sections of 
${\bf T}\to Z$, let $\cA_T\subset C^\infty(Z)$ denote the subalgebra of 
 functions $f$ whose Hamiltonian vector fields $u_f$ fulfill the condition
\mar{qq105}\beq
[u_f,\bT(Z)]\subset \bT(Z). \label{qq105}
\eeq
Elements of this subalgebra are quantized only. Metaplectic
correction provides the pre-Hilbert space $E_T$ where the quantum algebra
$\cA_T$ acts by symmetric operators. This is a certain subspace of sections of
the tensor product $C\ot\cD$ of the prequantization line bundle $C\to Z$ and a
bundle $\cD\to Z$ of half-densities on $Z$.  
The
geometric quantization  procedure has been extended to Poisson manifolds
\cite{vais91,vais} and to Jacobi manifolds
\cite{leon}.

Geometric quantization of the cotangent bundle $T^*Q$ is well-known
(e.g., \cite{eche98,sni,wood}). The
problem is that geometric quantization of 
$T^*Q$ does
not automatically imply quantization of non-relativistic mechanics. 

The momentum phase space of non-relativistic mechanics is the vertical
cotangent bundle
$V^*Q$ of $Q\to\bR$, 
endowed with the holonomic coordinates $(t=q^0,q^k,p_k)$.
It is provided with the canonical Poisson structure 
\mar{m72}\beq
\{f,f'\}_V = \dr^kf\dr_kf'-\dr_kf\dr^kf', \qquad f,f'\in C^\infty(V^*Q),
\label{m72}
\eeq
whose
symplectic foliation coincides with the fibration $V^*Q\to\bR$
\cite{book98,sard98}. 
Of course, one can 
quantize directly  the Poisson manifold $V^*Q$.
The problem is that non-relativistic mechanics can not be described as a
Poisson Hamiltonian system on the momentum phase space $V^*Q$. 
Its Hamiltonian $\cH$ is not an element of the Poisson algebra
$C^\infty(V^*Q)$, but a global section of the one-dimensional affine bundle
\mar{z11}\beq
\zeta:T^*Q\to V^*Q. \label{z11}
\eeq
Therefore, the non-relativistic Hamilton equation can be written as the
constraint Hamilton equation (\ref{qq91}) on the cotangent bundle $T^*Q$.

Thus, we need compatible geometric quantizations both of the
cotangent bundle $T^*Q$ and the vertical cotangent bundle $V^*Q$
\cite{eprint}. We use that the fibration (\ref{z11}) 
defines the symplectic
realization of the Poisson manifold $(V^*Q,\{,\}_V)$, i.e., 
\mar{qq8}\beq
\zeta^*\{f,f'\}_V=\{\zeta^*f,\zeta^*f'\}_T \label{qq8}
\eeq
for all $f,f'\in C^\infty(V^*Q)$. As a consequence, there is the 
monomorphism
\mar{qq60}\beq
\zeta^*: (C^\infty(V^*Q),\{,\}_V) \to (C^\infty(T^*Q),\{,\}_T) \label{qq60}
\eeq
of the Poisson algebra on $V^*Q$ to that on $T^*Q$. 

The
problem is that, though prequantization of $T^*Q$ leads to prequantization
of $V^*Q$, polarization of $T^*Q$ need not imply polarization of
the Poisson manifold $V^*Q$. We will show that the Schr\"odinger
representation of $T^*Q$  by operators on half-densities on $Q$ yields the
geometric quantization of $V^*Q$ such that the monomorphism of Poisson
algebras (\ref{qq60}) can be prolonged to that of quantum algebras.  This
prolongation is also required in order that non-relativistic and relativistic
geometric quantizations be compatible  on a local chart (\ref{gm600}) of the
relativistic configuration space $Q$. Such a compatibility takes place in
heuristic quantum theory where space-time coordinates
$q^\la$ and spatial momenta
$p_k$ have the same Schr\"odinger representation in non-relativistic and
relativistic quantum mechanics. Given the Schr\"odinger representation of
$T^*Q$, the quantum constraint equations
make a sense of the Schr\"odinger equation in non-relativistic quantum
mechanics and the relativistic quantum wave equation in the presence of a
background metric $g$.

A fault of the Schr\"odinger representation of $T^*Q$ is that the corresponding
quantum algebra
$\cA_T$ consists of functions which are at most affine in momenta. Therefore,
it does not include the most of physically relevant Hamiltonians. If a
Hamiltonian $\cH$ is a quadratic in momenta, one can represent it as an
element of the universal algebra of the Lie algebra $\cA_T$, but this
representation is not always globally defined. This problem 
becomes especially important if one considers
the non-relativistic limit of a relativistic systems  because a transitive
Hamiltonian is not a polynomial. It means that there is no satisfactory
transition between quantum non-relativistic and relativistic systems.

\section{Non-relativistic Hamiltonian dynamics}

The dynamic equation (\ref{qq91}a) is obtained as follows.
Every global section
\mar{qq4}\beq
h:V^*Q\to T^*Q, \qquad p\circ
h=-\cH(t,q^j,p_j) \label{qq4}
\eeq
of the affine bundle $\zeta$ (\ref{z11}) yields  the
pull-back Hamiltonian form
\mar{b4210}\beq
H=h^*\Xi= p_k dq^k -\cH dt  \label{b4210}
\eeq
on $V^*Q$. Given a trivialization 
\mar{gm156}\beq
V^*Q\cong\bR\times T^*M, \label{gm156}
\eeq
the form $H$ is the
well-known  integral invariant of Poincar\'e--Cartan, where $\cH$ is a
Hamiltonian \cite{arno}. 
Given a Hamiltonian form $H$ (\ref{b4210}), there exists a unique
Hamiltonian connection 
\mar{m57}\beq
\g_H=\dr_t +\dr^k\cH\dr_k -\dr_k\cH\dr^k \label{m57}
\eeq
on the fibre bundle $V^*Q\to \bR$ such that
\mar{qq1}\beq
\g_H\rfloor dH=0 \label{qq1}
\eeq
\cite{book98,sard98}. It defines the Hamilton equations on $V^*Q$.

Let us consider the pull-back
$\zeta^*H$ 
of the
Hamiltonian form
$H=h^*\Xi$ onto the cotangent bundle $T^*Q$. It is readily observed that the
difference
$\Xi-\zeta^*H$ is a horizontal 1-form on $T^*Q\to\bR$. Then the contraction  
\mar{mm16}\beq
\cH^*=\dr_t\rfloor(\Xi-\zeta^*H)=p+\cH \label{mm16}
\eeq
is a function on $T^*Q$. It is exactly the constraint function $\f_N$
(\ref{qq90}a) of the image $N=\im h$ of the closed imbedding $h$
(\ref{qq4}). It is given by the
constraint 
\mar{rq2}\beq
\cH^*=p+\cH(t,q^k,p_k)=0. \label{rq2}
\eeq
Let us consider the equation
\mar{rq1}\beq
\g\rfloor i_N^*\Om=0, \qquad i_N:N\to T^*Q, \label{rq1}
\eeq
for a vector field $\g$ on $N$. 
It defines a
conservative Dirac constraint system with a zero Hamiltonian on the primary
constraint subspace (\ref{qq90}a) of the symplectic manifold
$T^*Q$. Being a one-codimensional closed imbedded submanifold, the constraint
space $N$ is coisotropic. Therefore, a solution of the equation (\ref{rq1})
always exists \cite{book98,marmo}. If 
$\g\rfloor dt=1$, it is
readily observed that 
$T\zeta \g=\g_H$, where $T\zeta$
denotes the tangent morphism to $\zeta$ (\ref{z11}). 

A glance at the equation (\ref{qq1}) shows that one can think of the
Hamiltonian connection $\g_H$ as being the Hamiltonian vector field of a zero
Hamiltonian with respect to the presymplectic form $dH$ on $V^*Q$. 
Therefore, one can consider geometric quantization of the presymplectic
manifold
$(V^*Q,dH)$, besides geometric quantization of the Poisson manifold
$(V^*Q,\{,\})$. Given a
trivialization (\ref{gm156}), this quantization has been studied in
\cite{wood}.

Usually, geometric quantization is not applied directly
to a presymplectic manifold $(Z,\om)$, but to a symplectic manifold 
$(Z',\om')$
such that the presymplectic form $\om$ is a pull-back of the symplectic
form $\om'$. Such a symplectic manifold always exists. 
The following two possibilities are customarily considered: (i) 
$(Z',\om')$
is a reduction of $(Z,\om)$ along the leaves of the characteristic distribution
of the presymplectic form $\om$ of constant rank \cite{got86,vais83},
and (ii)
there is a coisotropic
imbedding of $(Z,\om)$ to $(Z',\om')$ \cite{got81,got82}.

In application to $(V^*Q,dH)$, the reduction procedure meets
difficulties as follows. Since the kernel of
$dH$ is generated by the vectors
($\dr_t$, $\dr_k\cH\dr^k-\dr^k\cH\dr_k$, $k=1,\ldots,m$),
the presymplectic form $dH$ in physical models is almost never of
constant rank. Therefore, one has to provide an
exclusive analysis of each physical model, and to 
cut out a certain subset of $V^*Q$
in order to use the reduction procedure.

The second variant of geometric quantization of the presymplectic manifold
$(V^*Q,dH)$ seems more attractive because  any section
$h$ (\ref{qq4}) is a coisotropic imbedding.  
Then the geometric quantization of the
presymplectic manifold
$(V^*Q,dH)$ consists in geometric quantization of the cotangent bundle
$T^*Q$ and setting the quantum constraint condition (\ref{qq92}a)
on physically admissible quantum states. 

Thus, presymplectic geometric
quantization of $V^*Q$  agrees with the  above manifested approach to
quantization of non-relativistic mechanics which requires additionally 
that geometric quantization of $T^*Q$ also provides quantization of the 
Poisson algebra on $V^*Q$. 

\section{Relativistic Hamiltonian dynamics}

We aim to show that classical Hamiltonian relativistic mechanics 
on the configuration space $Q$
can be seen as
a conservative
constraint Dirac system on the
cotangent bundle $T^*Q$. 

We start from describing the velocity and momentum  phase spaces of
relativistic mechanics \cite{book98,sard98}.

The velocity phase space of relativistic
mechanics is  the first order jet manifold 
$J^1_1Q$ of 1-dimensional submanifolds of the manifold $Q$. It consists of the
equivalence classes
 $[S]^1_q$, $q\in Q$, of 
one-dimensional imbedded submanifolds of $Q$ which pass through $q\in Q$ and
are tangent to each other at $q$. 
Given the coordinates 
$(q^0,q^k)$ on $Q$ with the transition functions
\mar{b5.1}\beq
q^0\to \wt q^0(q^0, q^j), \qquad q^k\to \wt q^k(q^0, q^j), \label{b5.1}
\eeq
the jet manifold $J^1_1Q$ is
endowed with the adapted coordinates 
$(q^0,q^k,q^k_0)$ whose transition functions are obtained as
follows. Let $d_0=\dr_0 + q^k_0\dr_k$
be the total derivative.
Given coordinate transformations (\ref{b5.1}), one can easily find that
\be
d_{\wt q^0}= d_{\wt q^0}(q^0)d_{q^0}=\left(\frac{\dr q^0}{\dr \wt q^0} +
\wt q^k_0\frac{\dr q^0}{\dr \wt q^k}\right)d_{q^0}.
\ee
Then we obtain the equation
\be
\wt q^k_0=  d_{\wt q^0}(q^0)d_{q^0}(\wt q^k) =
\left(\frac{\dr q^0}{\dr \wt q^0} +\wt q^k_0\frac{\dr q^0}{\dr \wt
q^k}\right) \left(\frac{\dr \wt q^k}{\dr q^0} +
q^j_0\frac{\dr \wt q^k}{\dr q^j}\right).
\ee
Its solution is
\mar{z799}\beq
\wt q^k_0= \left(\frac{\dr \wt q^k}{\dr q^0} +
 q^j_0\frac{\dr \wt q^k}{\dr q^j}\right)\mbox{\large /}
 \left(\frac{\dr \wt q^0}{\dr q^0} + q^k_0\frac{\dr \wt q^0}{\dr q^k}\right).
\label{z799}
\eeq
A glance at the transformation law (\ref{z799}) shows that 
the fibration $\pi:J^1_1Q\to Q$
is a projective bundle.

\begin{ex}
Put $Q=\bR^4$ whose 
Cartesian coordinates $(q^0,q^k)$ are
subject to the Lorentz transformations
\mar{gm605}\beq
\wt q^0= q^0{\rm ch}\al  -q^1{\rm sh}\al,\qquad
 \wt q^1=-q^0{\rm sh}\al + q^1{\rm ch}\al, \qquad
\wt q^{2,3} = q^{2,3}. \label{gm605}
\eeq
Then (\ref{z799}) is exactly the transformations
\be
\wt q^1_0=\frac{-{\rm sh}\al + q^1_0{\rm ch}\al}{{\rm ch}\al - q^1_0{\rm
sh}\al}, \qquad 
&& \wt q^{2,3}_0= \frac{q^{2,3}_0}{{\rm ch}\al - q^1_0{\rm sh}\al}
\ee
of three-velocities in Special Relativity.
\end{ex} 

Thus, one can think of the velocity phase space $J^1_1Q$ as being the space of
non-relativistic velocities of a relativistic system. 

The space of relativistic velocities is the tangent bundle
$TQ$ of $Q$ equipped with the holonomic
coordinates $(q^\la,\dot q^\la)$.
We have the map 
\mar{gm601}\beq
\la: J^1_1Q\ni q^k_0\mapsto (\dot q^0, \dot q^k=\dot q^0 q^k_0)\subset TQ,
\label{gm601}
\eeq
over $Q$ 
which assigns to each point of $J^1_1Q$ a line in $TQ$. There is the converse
map
\mar{b5.3}\beq
\vr: TQ \to J^1_1Q, \qquad q^k_0\circ \vr = \dot q^k/\dot q^0,
\label{b5.3}
\eeq
such that $\vr\circ\la =\id J^1_1Q$. It should be
emphasized that, though the expression (\ref{b5.3}) looks
singular at $\dot q^0=0$, this point belongs to another coordinate chart,
and the morphism
$\vr$ is well defined.

A pseudo-Riemannian metric
$g$ on $Q$ defines the subbundle of 
hyperboloids 
\mar{z931}\beq
W_g=\{\dot q^\la\in TQ\, :\, g_{\m\nu}(q)\dot q^\m\dot q^\nu=1\}
\label{z931}
\eeq
of $TQ$. Then, restricting $\vr$ (\ref{b5.3}) and the image of $\la$
(\ref{gm601}) to $W_g$, we obtain the familiar relations 
between non-relativistic  and  relativistic velocities in relativistic
mechanics. Since $Q$ is assumed to be time-oriented, the subbundle  of
hyperboloids $W_g$ is a disjoint union of the subbundles $W_g^+$ ($\dot
q^0>0$) and $W_g^-$ ($\dot
q^0<0$). Hereafter, we will consider only its connected component $W_g^+$.

\begin{rem}
Note that, in non-relativistic mechanics on a configuration bundle $Q\to\bR$,
relativistic velocities of a non-relativistic system live in the subbundle
$\dot q^0=1$ of
$TQ$. In particular, any non-relativistic dynamic equation on
$Q\to\bR$ is equivalent to a geodesic equation with respect
to a connection on the tangent bundle $TQ\to Q$ \cite{book98,geod}.
\end{rem}

The standard Lagrangian formalism fails to be appropriate to
relativistic mechanics in a straightforward manner because a Lagrangian 
\mar{z950}\beq
L=\cL(q^\la,q^k_0) dq^0, \label{z950}
\eeq
is defined only locally on a coordinate chart
$(U;q^0,q^k)$ of the velocity phase space
$J^1_1Q.$ Nevertheless, 
given a motion $q^k=c^k(q^0)$, the pull-back $c^*L$ of a 
Lagrangian (\ref{z950}) is well behaved under transformations (\ref{b5.1})
where $d\wt q^0=d_{q^0}(\wt q^0)dq^0$.

Therefore, let us consider a local regular Lagrangian $L$ (\ref{z950}) on a
coordinate chart
$(U;q^0,q^i,q^i_0)$ of $J^1_1Q$, treated as a
local velocity phase space of a non-relativistic mechanical system. This
system can be described as a local Dirac constraint system on the cotangent
bundle $T^*Q|_U$ in the framework of the well-known Hamilton--De Donder
formalism (e.g., \cite{book98}). Indeed, the  Poincar\'e--Cartan form
associated to the Lagrangian
$L$ defines the Legendre morphism
\mar{z960}\beq
\wh H_L:J^1Q|_U\to T^*Q|_U, \qquad 
p_0= \cL-q^i_0\frac{\dr\cL}{\dr q^i_0}, \qquad p_i=\frac{\dr\cL}{\dr
q^i_0}, \label{z960}
\eeq
where the cotangent bundle $T^*Q$ of
$Q$ plays a role of the homogeneous momentum phase space of
non-relativistic mechanics. 
If a Lagrangian $L$ is regular, the equations (\ref{z960}) are solvable
uniquely for
\mar{q1}\beq
p_0=\f(q^\m, p_i). \label{q1}
\eeq
Then a solution of the Lagrange equations is an integral
curve of the vector field $\g$ on the constraint space $N$ (\ref{q1}) which
fulfills the equation
\be
\g\rfloor \Om_N=0,
\ee
where $\Om_N$  is the pull-back onto $N$ of the symplectic
form $\Om$ on $T^*Q$. This is a Dirac
constraint system on $T^*Q$ in the case of a constant Hamiltonian on the
primary constraint space (\ref{q1}). 

For instance, the Lagrangian 
\mar{z940}\beq
L_m=-m(1-\op\sum_i(q^i_0)^2)^{1/2}dq^0 \label{z940}
\eeq
of a free relativistic point mass $m$ in Special Relativity is regular on the
ball $\op\sum_i(q^i_0)^2<1$,
and defines a Dirac constraint system with a constant Hamiltonian on the
primary constraint space
\mar{z961}\beq
p^2_0-\op\sum_i p_i^2=m^2. \label{z961}
\eeq

Therefore, let us describe a relativistic mechanical system as a conservative
Hamiltonian system on the symplectic manifold $T^*Q$ which is characterized by
a relativistic Hamiltonian 
\mar{gm609}\beq
\bH: T^*Q\to\bR \label{gm609}
\eeq 
\cite{book98,rov91,sard98}. One also considers $T^*Q$,
but provided with another symplectic form, e.g., $\Om + F$ where $F$ is the
strength of an electromagnetic field \cite{sni}. 

Any
relativistic Hamiltonian $\bH$ (\ref{gm609}) defines the Hamiltonian map 
\mar{gm616}\beq
\wh\bH: T^*Q\to TQ, \qquad \dot q^\m=\dr^\m\bH, \label{gm616}
\eeq
over $Q$ from the relativistic momentum phase space  $T^*Q$ to the
 space $TQ$ of relativistic velocities. Since the relativistic velocities of a
relativistic system live in the velocity hyperboloid $W_g^+$, we
have the constraint subspace $N'=\wh\bH^{-1}(W_g^+)$ (\ref{qq90}b)
of the relativistic momentum phase space $T^*Q$. Let us assume that $N'\to Q$
is a closed imbedded one-codimensional subbundle of $T^*Q$ and, consequently,
is coisotropic. It takes place if the Hamiltonian map (\ref{gm616}) is of
constant rank. Then a relativistic mechanical system can be described as a
conservative  Dirac constraint system on the primary constraint space
$i_{N'}:N'\to T^*X$ (\ref{qq90}b). Its solutions are
integral curves of the Hamiltonian vector field 
$\g'$
on
$N'$ which obeys the relativistic Hamilton equation
\mar{gm610}\beq
\g'\rfloor i^*_{N'}\Om=-i_{N'}^* d\bH. \label{gm610} 
\eeq
A simple calculation shows that, if 
\mar{rq10}\beq
\{\cH,g_{\m\nu}\dr^\m\cH\dr^\nu\cH\}_T=0, \label{rq10}
\eeq
the equation (\ref{gm610}) has a solution
\mar{rq11}\beq
\g'^\la=\dr^\la\cH, \qquad \g'_\la=-\dr_\la\cH. \label{rq11}
\eeq

Let us give a few basic examples of relativistic Hamiltonian systems.

\begin{ex} \label{gm621} \mar{gm621}
The relativistic Hamiltonian of a free relativistic point mass in Special
Relativity is
\mar{z973}\beq
\bH= -\frac1{2m}\eta^{\m\n}p_\m p_\nu, \label{z973}
\eeq
where $\eta$ is the Minkowski metric, while the constraint space
$N'$ is given by the equation (\ref{z961}). It is readily that this
Hamiltonian fulfills the condition (\ref{rq10}).  Moreover,  its restriction
 to the constraint space  is a constant
function. Then the Hamilton equation (\ref{gm610}) takes the form
\beq
u\rfloor i^*_N\Om=0. \label{gm613}
\eeq
Its solution (\ref{rq11}) reads
\be
&& \g'_\la=0, \qquad p_k={\rm const}, \qquad
p_0=-(m^2-\eta^{jk}p_jp_k)^{-1/2}, \\
&& \g'^\la= -\frac1{m}\eta^{\la\nu}p_\nu.
\ee
Then, we obtain the familiar expression for three-velocities
\be
q^i_0= -\eta^{il}p_l(m^2-\eta^{jk}p_jp_k)^{-1/2}. 
\ee
\end{ex}

\begin{ex} \label{gm622} \mar{gm622}
Let us consider a point electric charge $e$ in the Minkowski space in the
presence of an electromagnetic potential $A_\la$. Its relativistic
Hamiltonian reads
\be
\bH=-\frac1{2m} \eta^{\m\nu}(p_\m-eA_\m) (p_\nu-eA_\nu),  
\ee
while the constraint space $N'$ (\ref{qq90}b) is
\be
\eta^{\m\nu}(p_\m-eA_\m) (p_\nu-eA_\nu)=m^2.
\ee
As in the previous example, the Hamilton $\bH$ 
fulfills the condition (\ref{rq10}), and  its restriction
to the constraint space  is a constant
function. Therefore, the relativistic Hamilton
equation (\ref{gm610}) takes the
form (\ref{gm613}). Its solution (\ref{rq11}) is
\mar{gm613'4}\ben
&& \g'_\la=-\frac{e}{m}\eta^{\m\nu}\dr_\la A_\m(p_\nu-eA_\nu), \label{gm613'}
\\ 
&& \g'^i =-\frac1m \eta^{ik}(p_k- eA_k), \qquad 
\g'^0=\frac1m\eta^{00}(m^2-\eta^{ij}(p_i-eA_i)(p_j-eA_j))^{-1/2}.
\label{gm614}
\een
The equality (\ref{gm614}) leads to the usual expression for the
three-velocities
\be
p_k=-m\eta_{ki}q^i_0(1+\eta_{ij}q^i_0q^j_0)^{-1/2}+A_k
\ee
Substituting this expression in the equality (\ref{gm613'}), we obtain the
familiar equation of motion of a relativistic charge in an electromagnetic
field.
\end{ex}

\begin{ex} \label{gm623} \mar{gm623}
The relativistic Hamiltonian for a point mass $m$ in a
gravitational field $g$ on a 4-dimensional manifold $Q$ reads
\mar{rq20}\beq
\bH=-\frac1{2m}g^{\m\nu}(q)p_\m p_\nu, \label{rq20}
\eeq
while the constraint space $N'$ (\ref{qq90}b) is
\be
g^{\m\nu}p_\m p_\nu =m^2.
\ee
As in previous Examples, the relativistic Hamilton equation (\ref{gm610}) takes
the form (\ref{gm613}).
\end{ex}

\section{Prequantization}

Basing on the standard prequantization of the cotangent bundle $T^*Q$,
we will construct the compatible prequantizations of the Poisson
bundle $V^*Q\to\bR$. 

Let us recall the prequantization of $T^*Q$ (e.g., \cite{eche98,sni,wood}). 
Since its symplectic form $\Om$ is exact and, consequently, is of zero de
Rham cohomology class,  
the prequantization bundle is the trivial complex line bundle 
\mar{qm501}\beq
C=T^*Q\times\bC\to T^*Q,  \label{qm501}
\eeq
whose Chern class $c_1$ is zero. Coordinated by $(q^\la,p_\la,c)$,
it is provided with the
admissible linear connection
\mar{qm502}\beq
A=dp_\la\ot\dr^\la +dq^\la\ot(\dr_\la+ ip_\la c\dr_c) \label{qm502}
\eeq
with the strength form $F= - i\Om$ and the
Chern form 
\be
c_1=\frac{i}{2\pi}F=\frac1{2\pi}\Om.
\ee
The $A$-invariant Hermitian fibre metric on $C$ is
$g(c,c)=c\ol c$.
The covariant derivative of sections $s$ of the prequantization bundle $C$
(\ref{qm501}) relative to the connection $A$ (\ref{qm502}) along the vector
field
$u$ on $T^*Q$ takes the form
\mar{qq12}\beq
\nabla_u(s)= (u^\la\dr_\la - iu^\la p_\la)s. \label{qq12}
\eeq
Given a function $f\in C^\infty(T^*Q)$, 
the covariant derivative (\ref{qq12}) along 
the Hamiltonian vector field 
\be
u_f=\dr^\la f\dr_\la -\dr_\la f\dr^\la, \qquad u_f\rfloor\Om=-df,
\ee
of $f$ reads
\be
\nabla_{u_f}= \dr^\la f(\dr_\la - ip_\la) - \dr_\la f\dr^\la.
\ee
Then, in order to satisfy the Dirac condition (\ref{qm514}), one
assigns to each function
$f\in C^\infty(T^*Q)$ the first order differential
operator 
\mar{qm504}\beq
\wh f(s) =-i(\nabla_{u_f} + if)s=[-iu_f+ (f -p_\la\dr^\la f)]s
\label{qm504}
\eeq
on sections
$s$ of the prequantization bundle $C$ (\ref{qm501}).
For instance, the prequantum operators (\ref{qm504}) of local functions 
$f=p_\la$, $f=q^k$ and global functions $f=t$, $f= 1$  read
\be
\wh p_\la= -i\dr_\la, \qquad \wh q^\la =i \dr^\la +  q^\la,
\qquad \wh 1= 1. 
\ee
For elements $f$ of the Poisson subalgebra $\zeta^*C^\infty(V^*Q)\subset
C^\infty(T^*Q)$, the Kostant--Souriau
formula (\ref{qm504}) takes the form
\mar{qq40}\beq
\wh f(s) =[-i(\dr^kf\dr_k-\dr_\la f\dr^\la)+ (f -p_k\dr^k f)]s.
\label{qq40}
\eeq

Turn now to
prequantization of the Poisson manifold $(V^*Q,\{,\}_V)$.  
The Poisson bivector $w$ of the Poisson structure (\ref{m72}) on
$V^*Q$ reads
\mar{qq41}\beq
w=\dr^k\w\dr_k=-[w,u]_{\rm SN}, \label{qq41}
\eeq
where $[,]_{\rm SN}$ is the Schouten--Nijenhuis bracket and $u=p_k\dr^k$ is
the Liouville vector field on the vertical cotangent bundle $V^*Q\to Q$.
The relation (\ref{qq41}) shows that the Poisson bivector $w$ is
exact and, consequently, has the zero Lichnerowicz--Poisson cohomology
class
\cite{book98,vais}. Therefore, let us consider the trivial complex line bundle 
\mar{qq43}\beq
C_V=V^*Q\times \bC\to V^*Q \label{qq43}
\eeq
such that the zero Lichnerowicz--Poisson cohomology class of $w$ is the image
of the zero Chern class $c_1$ of $C_V$ under the cohomology homomorphisms 
\be
H^*(V^*Q,\bZ) \to H^*_{\rm deRh}(V^*Q) \to H^*_{\rm LP}(V^*Q).
\ee

Since the line bundles $C$ (\ref{qm501}) and $C_V$ (\ref{qq43}) are trivial, 
$C$
can be seen as the pull-back
$\zeta^*C_V$ of $C_V$, while $C_V$ is isomorphic to the
pull-back
$h^*C$ of $C$ with respect to a section 
$h$ (\ref{qq4}) of the affine bundle (\ref{z11}).   
Since $C_V=h^*C$ and since the covariant derivative of the connection $A$
(\ref{qm502}) along the fibres of $\zeta$ (\ref{z11}) is trivial, let us
consider the pull-back 
\mar{qq44}\beq
h^*A=dp_k\ot\dr^k +dq^k\ot(\dr_k+ ip_k c\dr_c)+ 
dt\ot(\dr_t - i\cH c\dr_c) \label{qq44}
\eeq
of the connection $A$ (\ref{qm502}) onto $C_V\to V^*Q$
\cite{book00}.
This connection defines the
contravariant derivative 
\mar{qq45}\beq
\nabla_\f s_V = \nabla_{w^\sharp\f}s_V \label{qq45}
\eeq
of sections $s_V$ of $C_V\to V^*Q$ along one-forms $\f$ on $V^*Q$. This
contravariant derivative is corresponded to a contravariant connection $A_V$
on the line bundle
$C_V\to V^*Q$ \cite{vais}. It is readily observed that this
contravariant connection does not depend on the choice of a section $h$. 
By virtue of the relation (\ref{qq45}), the curvature bivector of $A_V$
is equal to
$-iw$  \cite{vais97}, i.e., $A_V$ is an admissible connection
for the canonical Poisson structure on $V^*Q$. Then the 
 Kostant--Souriau formula 
\mar{qq46}\beq
\wh f_V(s_V) =(-i\nabla_{u_{Vf}}+f)s_V=[-i(\dr^kf\dr_k-\dr_kf\dr^k)+ (f
-p_k\dr^k f)]s_V
\label{qq46}
\eeq
defines prequantization of the Poisson manifold $V^*Q$, where $s_V$ are
sections of the line bundle $C_V$. 

It is immediately observed that the prequantum operator $\wh f_V$
(\ref{qq46}) coincides with the prequantum operator $\wh{\zeta^*f}$
(\ref{qq40}) restricted to the pull-back sections $s=\zeta^*s_V$ of the line
bundle $C$. Thus, prequantization of the Poisson algebra $C^\infty(V^*Q)$ on
the Poisson manifold $(V^*Q,\{,\})$ is equivalent to its prequantization as a
subalgebra of the Poisson algebra $C^\infty(T^*Q)$ on the symplectic 
manifold $T^*Q$.

\section{Polarization}

Given compatible prequantizations of the cotangent bundle $T^*Q$ and the
Poisson bundle $V^*Q\to \bR$, let us now construct
their compatible polarizations.  

Recall that, given a polarization $\bT$ of a prequantum symplectic manifold
$(Z,\Om)$, the subalgebra $\cA_T\subset C^\infty(Z)$ of 
functions $f$ obeying the condition
(\ref{qq105}) is only quantized.
Moreover, after further metaplectic correction, one considers
a representation of this algebra in a quantum space $E_T$ such
that
\mar{qq15}\beq
\nabla_u e =0, \qquad \forall u\in \bT(Z), \qquad e\in E_T. \label{qq15}
\eeq

Recall that a polarization of a Poisson manifold $(Z,\{,\})$ is defined as a
sheaf
$\bT^*$ of germs of complex functions on $Z$ whose stalks $\bT^*_z$, $z\in Z$,
are Abelian algebras with respect to the Poisson bracket $\{,\}$ \cite{vais97}.
One can also require that the algebras $\bT^*_z$ are maximal,
but this condition need not hold under pull-back and push-forward operations. 
Let $\bT^*(Z)$ be the structure
algebra  of global sections of the sheaf $\bT^*$; it is
also called a Poisson polarization \cite{vais91,vais}.
A quantum algebra $\cA_T$ associated to the Poisson polarization $\bT^*$ is
defined as a subalgebra of the Poisson algebra
$C^\infty(Z)$ which consists of functions $f$ such that
\be
\{f,\bT^*(Z)\}\subset \bT^*(Z).
\ee
Polarization of a symplectic manifold yields its maximal Poisson
polarization, and {\it vice versa}.

There are different polarizations of the cotangent bundle $T^*Q$. We will
consider those of them whose direct images as Poisson
polarizations  onto
$V^*Q$ with respect to the morphism $\zeta$ (\ref{z11}) are polarizations of
the Poisson manifold $V^*Q$. This takes place if the germs of
a polarization 
$\bT^*$ of
the Poisson manifold $(T^*Q,\{,\}_T)$ are constant along the fibres of the
fibration $\zeta$ (\ref{z11}) \cite{vais97}, i.e., are germs of functions
independent of the momentum coordinate $p_0=p$. It means that the corresponding
symplectic polarization
$\bT$ of 
$T^*Q$ is vertical with respect to the fibration $T^*Q\to\bR$.

The vertical polarization $VT^*Q$ of $T^*Q$ obeys this
condition. It
is a strongly admissible polarization, and its integral manifolds are fibres of
the cotangent bundle
$T^*Q\to Q$.
One can easily verify that the associated quantum
algebra $\cA_T$ consists of functions on $T^*Q$ which are at most affine in
momenta
$p_\la$. 
The quantum space $E_T$ associated to the vertical polarization obeys the
condition 
\mar{qq80}\beq
\nabla_{u^\la\dr_\la} e=0, \qquad \forall e\in E_T. \label{qq80}
\eeq
Therefore, the operators of the quantum algebra $\cA_T$ on this quantum space
read
\mar{qq20}\beq
f=a^\la(q^\mu)p_\la + b(q^\mu), \qquad \wh f= -i\nabla_{a^\la\dr_\la} +b.
\label{qq20}
\eeq
This is the Schr\"odinger representation of $T^*Q$.

The vertical polarization of $T^*Q$ defines the maximal polarization
$\bT^*$ of the Poisson manifold $V^*Q$ which consists of germs of functions,
constant on the fibres of $V^*Q\to Q$. The
associated quantum space $E_V$ obeys the condition 
\mar{qq70}\beq
\nabla_{u^k\dr_k} e=0, \qquad \forall e\in E_V. \label{qq70}
\eeq
The quantum algebra $\cA_V$ corresponding to this polarization of
$V^*Q$ consists of functions on $V^*Q$ which are at most affine in momenta
$p_k$. Their quantum operators read
\mar{qq20'}\beq
f=a^k(q^\mu)p_k + b(q^\mu), \qquad \wh f= -i\nabla_{a^k\dr_k} +b.
\label{qq20'}
\eeq
This is the Schr\"odinger representation of $V^*Q$.

\section{Metaplectic correction}

To complete the geometric quantization procedure of $V^*Q$,
let us consider the metaplectic correction of the
Schr\"odinger representations of $T^*Q$ and
$V^*Q$.

The representation of the quantum algebra $\cA_T$ (\ref{qq20}) can be
defined in the subspace of sections of the line bundle $C\to T^*Q$ which
fulfill the relation (\ref{qq80}). 
This representation reads
\mar{qq81}\beq
f=a^\la(q^\mu)p_\la + b(q^\mu), \qquad \wh f= -ia^\la\dr_\la +b,
\label{qq81}
\eeq
and, therefore, can be restricted to sections of the pull-back
line bundle
$C_Q=\wh 0^*C\to Q$, where $\wh 0$ is the canonical zero section of
the cotangent bundle $T^*Q\to Q$.
However, this is not yet a representation
in a Hilbert space. 

Let $Q$ be an oriented manifold. Applying the general
metaplectic technique \cite{eche98,wood}, we come to the
vector bundle $\cD_{1/2}\to Q$ of complex half-densities on $Q$ with
the transition functions $\rho'=J^{-1/2}\rho$,
where $J$ is the Jacobian of the coordinate transition functions 
on $Q$. Since $C_Q\to Q$ is a trivial bundle, the tensor product
$C_Q\ot\cD_{1/2}$ is isomorphic to $\cD_{1/2}$. Therefore, the quantization
formula (\ref{qq81}) can be extended to sections of the half-density bundle
$\cD_{1/2}\to Q$ as 
\mar{qq82}\beq
f=a^\la(q^\mu)p_\la + b(q^\mu), \qquad
\wh f\rho=(-i\bL_{a^\la\dr_\la} +b)\rho= (-ia^\la\dr_\la 
-\frac{i}{2}\dr_\la (a^\la)+ b)
\rho,
\label{qq82}
\eeq
where $\bL$ denotes the Lie derivative.
The second term in the right-hand side of
this formula is a  metaplectic correction. It makes the operator $\wh f$
(\ref{qq82})  symmetric with respect to the Hermitian form
\be
\lng \rho_1|\rho_2\rng=\left(\frac{1}{2\pi}\right)^m
\op\int_Q \rho_1 \ol\rho_2
\ee
on the pre-Hilbert space $E_T$ of sections of
$\cD_{1/2}\to Q$ of compact support. The completion $\ol E_T$ of $E_T$ provides
a Hilbert space of the
Schr\"odinger representation of the quantum algebra $\cA_T$, where the
operators (\ref{qq82}) are essentially
self-adjoint. Of course, functions of
compact support on the time axis $\bR$ have a limited physical application,
but we can simply restrict our consideration to some bounded interval of
$\bR$. 

Since, in the case of the vertical polarization, there is a monomorphism of
the quantum algebra $\cA_V$ to the quantum algebra $\cA_T$, one can define
the  Schr\"odinger representation of $\cA_V$ by the operators
\mar{qq83}\beq
f=a^k(q^\mu)p_k + b(q^\mu), \qquad
\wh f\rho= (-ia^k\dr_k -\frac{i}{2}\dr_k(a^k) +b) \rho
\label{qq83}
\eeq
in the same space of complex half-densities on $Q$ as that of $\cA_T$.

\section{Quantum relativistic Hamiltonians}

The representation (\ref{qq83}) can be extended locally to functions on
$T^*Q$ which are polynomials of momenta $p_\la$ if they are
represented by elements of the universal enveloping algebra $\ol\cA_T$ of the
Lie algebra $\cA_T$. However, this representation is not necessarily
globally defined. For instance, a generic relativistic quadratic Hamiltonian 
\mar{qq130}\beq
\cH'=a^{\al\bt}(q^\la)p_\al p_\bt + b^\al(q^\la)p_\al + c(q^\la) \label{qq130}
\eeq
is not an element of the enveloping algebra
$\ol\cA_T$ because of the quadratic term.  Written 
locally as a Hermitian element $\wh p_\al \wh a^{\al\bt} \wh p_\bt$ of
$\ol\cA_T$, this term is quantized as 
\be
-\dr_\al(a^{\al\bt}\dr_\bt\rho)
\ee
 where
the derivative of half-density $\dr_\bt\rho$ is ill-behaved, unless 
the Jacobian of the coordinate transition functions 
on $Q$ is independent of fibre coordinates $q^k$ on $Q$. In particular, this
is the case of Special Relativity. For instance, the quantum relativistic
Hamiltonian of a free relativistic point mass in the Minkowski space reads
\be
\wh\bH= \frac{1}{2m} \eta^{\mu\nu}\dr_\mu\dr_\nu.
\ee
This is exactly the Laplace operator. Accordingly, the relativistic
Hamiltonian of a relativistic point electric charge takes the form
\be
\wh\bH= \frac{1}{2m} \eta^{\mu\nu}(\dr_\mu -i eA_\m)(\dr_\nu- ieA_\nu).
\ee
If the determinant of a pseudo-Riemannian metric $g$ on $Q$ 
is not constant, a rather sophisticated procedure of quantization of the
Hamiltonian (\ref{rq20}) has been
suggested in 
\cite{sni}. However, the Laplace operator constructed in
\cite{sni} does not fulfill the Dirac condition 
\be
[\wh \cH,\wh f]=-i\wh{\{\cH,f\}}_T, \qquad f\in\cA_T. 
\ee

Given a non-relativistic Hamiltonian $\cH$ and its Schr\"odinger quantization 
$\wh \cH$, the quantum constraint equation (\ref{qq92}a) is exactly the
Schr\"odinger equation
\be
i\dr_t \psi=\wh\cH\psi
\ee
of non-relativistic quantum mechanics.

A short calculation shows that, in the case of a free point mass and a point
relativistic charge, the quantum constraint equation (\ref{qq92}b) leads 
to the well-known equations of a classical scalar field. Their
interpretation as equations of relativistic quantum mechanics however is
under discussion. One of the problems is that the procedure of
the quantum non-relativistic limit fails to be well defined. For instance,
the Schr\"odinger quantization (\ref{qq92}b) of the classical constraint
(\ref{z961}) for a free relativistic point mass in the Minkowski space reads
\mar{rq31}\beq
(\eta^{\mu\nu}\dr_\mu\dr_\nu + m^2)\psi=0. \label{rq31}
\eeq
The classical constraint can be rewritten in an equivalent form 
\mar{rq30}\beq
p_0 \pm (m^2-\eta^{kj}p_kp_j)^{-1/2}=0, \label{rq30}
\eeq
suitable for passing to the non-relativistic limit
\be
p_0 \approx m -\frac{1}{2m}\eta^{kj}p_kp_j.
\ee
However, in contrast with the operator (\ref{rq31}), the operator
(\ref{rq30}) is not polynomial in momenta and its 
Schr\"odinger quantization is not defined even locally.

\end{document}